\documentclass[a4paper,fleqn]{cas-dc}

\usepackage[sort&compress,numbers]{natbib}
\def\tsc#1{\csdef{#1}{\textsc{\lowercase{#1}}\xspace}}
\tsc{WGM}
\tsc{QE}
\tsc{EP}
\tsc{PMS}
\tsc{BEC}
\tsc{DE}

\newcommand{\cz}{\textcolor{black}}
\newcommand{\eb}{\textcolor{black}}
\newcommand{\onlinecite}[1]{\hspace{-1 ex} \nocite{#1}\citenum{#1}}

\begin{document}
\let\WriteBookmarks\relax
\def\floatpagepagefraction{1}
\def\textpagefraction{.001}

\shorttitle{Recent use of ML and AI in crystal nucleation}

\shortauthors{Beyerle, Zou, and Tiwary}

\title [mode = title]{Recent advances in describing and driving crystal nucleation using machine learning and artificial intelligence}                      
% \tnotemark[1,2]

%\tnotetext[1]{This work is entirely funded by the US Department of Energy, Ofﬁce of Science, Basic Energy Sciences, CPIMS Program, under Award DE-SC0021009.}

%\tnotetext[2]{The second title footnote which is a longer text matter
%   to fill through the whole text width and overflow into
%   another line in the footnotes area of the first page.}

\author[1]{Eric R. Beyerle}[orcid=0000-0001-5717-3197]
\cormark[1]
\ead{ebeyerle@umd.edu}
%\cormark[1]



%\credit{Conceptualization of this study, Methodology, Software}

\affiliation[1]{organization={Institute for Physical Science and Technology},
    addressline={University of Maryland}, 
    city={College Park},
    % citysep={}, % Uncomment if no comma needed between city and postcode
    postcode={MD 20742}, 
    %state={MD},
    country={United States}
    }

\author[2]{Ziyue Zou}[]

\author[1,2]{Pratyush Tiwary}[]
%\ead{ptiwary@umd.edu}

%\credit{Data curation, Writing - Original draft preparation}

\affiliation[2]{organization={Department of Chemistry and Biochemistry},
    addressline={University of Maryland}, 
    city={College Park},
    % citysep={}, % Uncomment if no comma needed between city and postcode
    postcode={MD 20742}, 
    %state={MD},
    country={United States}
    }

% \affiliation[3]{organization={Institute for Physical Science and Technology},
%     addressline={University of Maryland}, 
%     city={College Park},
%     % citysep={}, % Uncomment if no comma needed between city and postcode
%     postcode={MD 20740}, 
%     %state={MD},
%     country={United States}
%     }

%\cortext[cor2]{Principal corresponding author}

%\fntext[fn1]{This is the first author footnote. but is common to third
%  author as well.}
%\fntext[fn2]{Another author footnote, this is a very long footnote and
%  it should be a really long footnote. But this footnote is not yet
%  sufficiently long enough to make two lines of footnote text.}

%\nonumnote{
%  }

\begin{abstract}
With the advent of faster computer processors and especially graphics processing units (GPUs) over the last few decades, the use of data-intensive machine learning (ML) and artificial intelligence (AI) has increased greatly, and the study of crystal nucleation has been one of the beneficiaries. In this review, we outline how ML and AI have been applied to address four outstanding difficulties of crystal nucleation: how to discover better reaction coordinates (RCs) for describing accurately non-classical nucleation situations; the development of more accurate force fields for describing the nucleation of multiple polymorphs or phases for a single system; more robust identification methods for determining crystal phases and structures; and as a method to yield improved course-grained models for studying nucleation.
\end{abstract}

% \includegraphics{figs/grabs.pdf}

%\item Research highlights item 1
%\item Research highlights item 2
%\item Research highlights item 3

% Keywords

\maketitle

\section{Introduction}
\label{sec:intro}
% Need to add a lot of citations to the introduction
For many years, the picture painted by classical nucleation theory (CNT)\cite{Sethna2021,deYoreo2020,Karthika2016,Peters2017} was seen as an accurate method for describing nucleation processes, and, indeed, the robustness of this method can be seen in its continued use for describing both the thermodynamics and kinetics of many nucleation processes.\cite{Sleutel2014,Parks2017,Duff2011} However, it has also become clear that not all nucleation events follow a purely classical pathway where the necessary and sufficient reaction coordinate (RC) is the size of the spherical crystal nucleus.\cite{piaggi2022icenucleation,LaCour2022Tuning,bertolazzo2022zeolite,Jacobson2010,Shtukenberg2019paracetamol,Karthika2016,Giberti2013, Finney2022naclnucleation, tsai2019reaction,Salvalaglio2012uncovering, Niu2019IceNucleation,gobbo2018infoS} Furthermore, even if $N$, \eb{the size of the nascent crystal nucleus,} can be shown to be a sufficiently good RC, CNT also stipulates that there is a single energy barrier along the RC corresponding to the barrier required to form the surface of the spherical critical nucleus of size $N^{*}$. However, it is well-known that for many nucleation processes the Ostwald step mechanism,\cite{vanSanten1984} whereby the most kinetically accessible crystalline structure forms first, followed by the more thermodynamically stable one, is the preferred nucleation pathway. This mechanism that requires the presence of at least two energetic barriers, one to go from the liquid state to the kinetically stabilized crystal structure and a second to go from the kinetically stabilized crystal structure to the thermodynamically stable crystal structure. 

With these two points in mind, it is clear, both from a simple thought experiment and from reported results, that, in at least some cases, CNT is not an effective way to describe the nucleation process. Unfortunately, at least some of the appeal of CNT is in its simplicity: it is an analytical, one-dimensional theory with both past and contemporary success. So, to obtain new RCs for the description of non-classical nucleation events, we must almost certainly use more nuanced, potentially non-linear RCs that are functions of many input features, which are {themselves} functions of the original coordinate space. One efficient method to construct such non-linear, feature-heavy RCs, {at least in theory}, is through the use of machine learning techniques, such as deep neural networks (deep NNs) of various flavors.\cite{Bonati2021Deep,Wang2021SPIB,ribeiro2018achieving,Varolgunecs2020,wang2020machine,wang2019past,mardt2018vampnets,karmakar2021cv,Sultan2018,Chen2018} These deep NNs are ideal in that they are, under certain conditions, universal function approximators,\cite{Hornik1989} meaning they should, with proper parameterization, be able to \eb{characterize} arbitrarily complex RCs for describing almost any nucleation process.\cite{Neha2023}
%\cz{should we also mention other problems in nucleation besides non-classical pathway to complete the picture in crystal nucleation bcuz i think non-classical pathway is important for NN RC development. ML on CG models may be helpful in facilitating crossing the high energy barrier in xtal nucleation}
%\eb{I agree with this comment. I think in the introduction, we need to do a better job of motivating \emph{why} we are taking the time to review what we are reviewing (i.e. why have we chosen to focus on using ML for finding RCs, force fields, polymorph identification methods, and new varieties of coarse graining?) Right now we have only really motivated why we are using ML to find novel RCs for nucleation processes, and we also need to motivate why we use ML for the other three categories.}

\eb{However, finding good RCs is just one component of analyzing nucleation processes. We must also have efficient and accurate techniques to classify both the initial isotropic phase and the final crystal structure. Furthermore, many systems of interest can crystallize into multiple polymorphs, with ice being one notable example.\cite{Brukhno2008ice,Bore2022}}
%\eb{{The use of machine learning (ML) techniques can also help in other areas in the study of nucleation. For example, a common problem in computer simulations of nucleation is deciding how to classify each frame of a trajectory into either a melt or a specific crystal structure.}} 
Many classical methods, \eb{i.e. those not influenced by ML,} for the identification of crystal structures have been developed \cite{Honeycutt1987, Larsen2016, Lazar2015,Nguyen2015,Ackland2006} and implemented \cite{Stukowski2010ovito, Ramasubramani2020} over time. Unfortunately, these methods can typically only sort structures into well-known crystal structures, such as hcp, fcc, bcc, sc, etc., and are not able to extrapolate well to systems at surfaces or with defects. The flexibility of ML and NNs allow for an increased expressiveness from a well-trained classifier, which can be more robust at classifying structures that are not perfect, canonical versions of these enumerated crystalline motifs.

%Likewise, the increased expressiveness offered by NNs and their relatives allows for the generation of force fields with density functional theory (DFT) level accuracy, but that can be evaluated with vastly superior performance compared to pure DFT.
\eb{From a computational perspective, to generate these crystal structures and the RCs necessary to describe how they are sampled, accurate and efficient force fields are required. The ability of deep NNs to successfully model complicated and highly non-linear functions, such as the energy surface of many physical systems, has spurred }their use in the development of accurate force fields for modeling nucleation via simulation. These NNs are more expressive compared to traditional \eb{molecular mechanics} force fields with their typically hand-tuned interaction parameters. Using ML to learn the energies for the configuration of a given system permits the generation of force fields with density functional theory (DFT) level accuracy, but allowing for simulations at speeds vastly superior to DFT level calculations.
For many nucleating systems, which include metals and semi-metals (e.g. gallium arsenide and silicon), quantum effects are important, and classical force fields cannot be used to simulate nucleation. 

\eb{Finally, at the microscopic scale studied in computer simulations, nucleation is a rare event occurring over typical simulation timescales. Even with accurate force fields, whether built using ML methods or not, nucleation events are difficult to sample computationally without using an enhanced sampling method or applying a biasing force to the nucleation RCs.\cite{Zou2023,Samanta2014,Schneider2017,Bonati2021Deep,giberti2015crystalnucleation} An alternative, fundamental approach to accelerate nucleation events is through an effective coarse-graining of the nucleating system. While many first-principles approaches for a systematic coarse-graining exist,\cite{zwanzig_book,Bird1987,Ruhle2009,McCarty2014} recently ML approaches have been utilized to find effective coarse-grains for nucleation, most notably water. The removal of fast molecular degrees of freedom irrelevant to the nucleation process allows for a larger integration timestep to be used, which accelerates the sampling of the system's relevant, coarse-grained phase space. This acceleration allows for enough statistics of the nucleation event to be collected to accurately study and describe the nucleation mechanism.} %ML techniques can also help to solve the rare event sampling problem of nucleation by using them to find optimized parameters for coarse-grained (CG) models of nucleating species, in particular water.

\begin{figure*}
  \centering
  \includegraphics[width=0.8\textwidth]{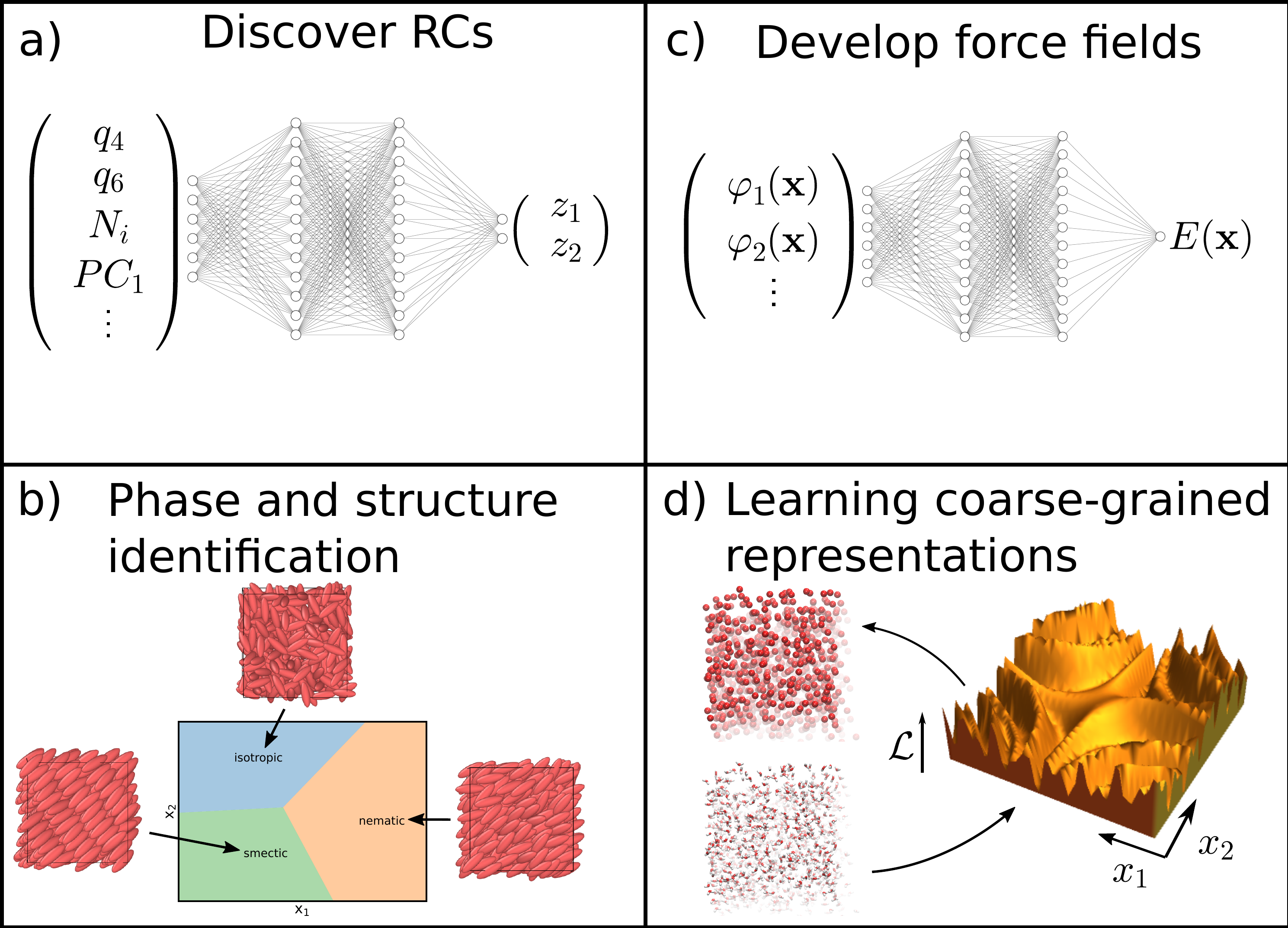}
  \caption
  {The applications of machine learning to nucleation of interest in this review. a) Neural networks and other machine learning architectures can be utilized to discover low-dimensional latent spaces where the important, slow processes describing nucleation can be described accurately and succinctly as non-linear combinations of the input features (Steinhardt bond order parameters; coordination numbers; principal components;$\ldots$); details given in Sec. \ref{RCs} b) Machine learning algorithms and various neural network architectures can be used to classify phases and polymorphs of a given condensed, atomic, or molecular system; details given in Sec. \ref{ID} c) The atomic coordinates $\mathbf{x}$ of an input structure can be transformed into a set of basis functions $\varphi_i(\mathbf{x})$. A neural network can be trained with these basis functions as input to return an energy of any given atomic configuration E($\mathbf{X}$); details given in Sec. \ref{FFs} d) Machine learning algorithms can optimize hyperparameters $x_1,x_2,\ldots$ with respect to a given loss function $\mathcal{L}$ over a high-dimensional hypersurface to generate a novel and accurate coarse-graining of an atomic or molecular system; details given in Sec. \ref{CG}. 
  }
  \label{fig:figure}
\end{figure*}

In this review, we will focus on \eb{these four distinct applications of ML and artificial intelligence (AI) to studies of nucleation: finding RCs, identifying crystal structures, developing force fields, and coarse-graining} (Figure \ref{fig:figure}). In Section \ref{RCs} we review how ML has allowed for the robust development of RCs going beyond the scope of CNT (Figure \ref{fig:figure}a). In Section \ref{ID}, we review how convolutional and graph neural networks have allowed for automatic phase and polymorph identification methods meeting or exceeding the benchmark performance delivered by established protocols such as polyhedral template matching and common neighbor analysis (Figure \ref{fig:figure}b). In Section \ref{FFs}, we review how neural networks have allowed for the impressive interpolation of force fields (FFs) rivalling DFT accuracy but allow for extended timescale simulations of metals and other electron-heavy systems to sample long timescale nucleation events (Figure \ref{fig:figure}c). In Section \ref{CG}, we review how ML has allowed for novel coarse-graining approaches to accelerate the sampling of nucleation events in a number of systems (Figure \ref{fig:figure}d). Finally, we conclude with a summary of current progress and speculate regarding ML and AI's future in the environment of nucleation research in Section \ref{future}. 

As a disclaimer, the usual caveat for review articles applies here: ML applications are rising rapidly in the field of nucleation, and we cannot possibly summarize and advertise all the studies performed. Thus, our choices are fundamentally biased by our own specific interests in this field, and the exclusion of any studies is due to kinetic constraints on our own ability to write a sufficiently detailed and comprehensive review article.

\section{Machine Learning for Discovering Reaction Coordinates Describing Nucleation}
\label{RCs}
\eb{In this section, we introduce general methods for discovering reactions coordinates and how machine learning offers techniques to automate RC discovery and tune the discovered RCs to the desired behavior (Figure \ref{fig:figure}a) We describe both shallow and deep ML approaches for RC discovery and generation.}
\subsection{What Generally Makes for a Good Reaction Coordinate?}
Most physical systems, from the quantum to the macroscopic, encompass processes that occur over a range of time- and lengthscales.\cite{vankampen2011,zwanzig_book} Often, the processes of interest are also some of the slowest processes of a system;\cite{Noe2012,Tiwary2016} for nucleation, this is also the case, as, typically, the nucleation and subsequent crystal growth processes tend to be some of the slowest microscopic processes a system can undergo. However, for nucleating systems, there \eb{will likely be} many irrelevant degrees of freedom, which might not contribute to the nucleation process. Thus, one major goal of nucleation research, from the microscopic perspective, is to find good RCs that are functions of the input variables that describe the nucleation process, which should exist in a subspace orthogonal to the variables irrelevant for nucleation to occur.

One of the standard methods for evaluating the goodness of an RC is through its comparison to the committor function, which describes the probability of entering next a pre-defined product state given a location in the input feature space instead of a pre-defined reactant state.\cite{Peters2017,Rhee2005,Chodera2011,Ma2005,E2010} The committor has been described as the ideal reaction coordinate because it tells the normalized distance between the reactant and product states, which is exactly what is desired in an RC. However, the weakness of the committor approach is that it requires \emph{a priori} information regarding the transformation of interest; if a slow degree of freedom is projected out prior to the committor calculation, then the resulting committor distribution will be deceptive and untrustworthy.\cite{Peters2017} Typical classical method for committor estimation include Markov state models \cite{Noe2009}, milestoning \cite{Elber2017}, and the string method. \cite{Ren2005}

Given that an appropriately constructed deep NN can serve as a universal function approximator, deep NNs are ideal tools to discover good RCs for describing nucleation, and recently much research in the field of nucleation has focused on developing various flavors of deep NNs for this task. One popular example is the use of an deep NN architecture known as an autoencoder,\cite{Bengio2013} which pushes the input variables through one or more layers of a fully connected NN before reaching the bottleneck layer, which encodes the input variables into a lower dimensional latent space. This latent space can then be decoded by pushing it through one or more layers of another NN called a decoder. The efficacy of the traditional autoencoder is based on the lower dimensional latent space's ability to reconstruct the input space with the minimum mean-squared error as the loss function. Since its inception, the autoencoder framework has been extended to encompass different loss functions.\cite{Lemke2019,Bengio2013,Wehmeyer2018,Kingma2014,Bakarji2022,Higgins2017} One important extension has been the introduction of a variational flavor where the evidence lower bound (ELBO)\cite{alemi2016deepVIB,Kingma2014} is minimized instead of the mean-squared error. Taking this approach allows for the construction of a low-dimensional RC that is maximally informative regarding the \eb{structures, though not necessarily dynamics,} of interest while maximally efficient at squeezing out the information regarding irrelevant processes.

In the absence of committor information, another successful approach is taking a variational approach, where the low-dimensional RC space is selected by applying a well-defined transformation to the input features and selecting the transformed features that maximize or minimize a given quantity. Perhaps the most well-known and heavily applied example of variational approaches is principal component analysis (PCA), where a linear transformation of the input features is found such that the first $n$ principle components defines an $n$-dimensional space that minimizes the mean-squared reconstruction error of the original, higher dimensional $N>n$ space.\cite{Jolliffe2002} However, it \eb{has been shown} that PCA can fail to find the most important processes in a system,\cite{perez2013identification,Beyerle2021a,Husic2016} so some studies have turned to the use of maximizing the timescales of the low-dimensional RC space through either linear methods such as time-lagged independent component analysis (tICA)\cite{Schwantes2013,perez2013identification,Takano1995} or through non-linear approaches such as the variational approach to Markov processes (VAMP)\cite{Nuske2014,Scherer2019} and the more general variational approach to conformational dynamics (VAC)\cite{Klus2018}.

\subsection{Machine Learning for Reaction Coordinate Generation}
Since ML approaches, perhaps most significantly deep NNs, are close to ideal for finding non-linear functions approximations, researchers in the field of RC discovery have begun to utilize NNs to find rich basis sets for the generation of RCs, especially those satisfying variational constraints.\cite{mardt2018vampnets,sidky2020molecular,Chen2019a,Bonati2021Deep} Especially popular has been the use of the autoencoder framework \cite{Bishop1995} with the loss function augmented to optimally reconstruct the either the input features\cite{Wehmeyer2018,ribeiro2018reweighted}, autocorrelation function\cite{Chen2019b}, or \eb{metastable state membership}  after a specified lagtime.\cite{Kube2007,Wang2021SPIB} Given the rising importance of ML approaches for contemporary RC generation, we outline in this section some methods used to discover good RCs for the nucleation.
\subsubsection{Autoencoders}
Multiple groups interested in studying the slow dynamics of various condensed matter systems have added a time lag to the autoencoder framework, yielding variational time-lagged autoencoder.\cite{Wehmeyer2018,sidky2020molecular} One specific type of time-lagged autoencoder, called the reweighted autoencoder variational Bayes (RAVE) method,\cite{ribeiro2018reweighted,Wang2021SPIB} has been used successfully to describe the nucleation of multiple polymorphs of urea from both the melt and aqueous solution, in addition to the nucleation of all three common polymorphs of glycine from aqueous solution.\cite{Zou2023} This approach is combined with metadynamics to provide a powerful method to both discover good RCs for nucleation and then enhance the sampling along those relevant, slow coordinates for nucleation to obtain good sampling for the liquid to polymorph transitions.

Another autoencoder architecture successfully developed to approximate the committor function has been developed by the Bolhuis group and their collaborators. They show in \cite{frassek2021extended} that the use of an autoencoder augmented with an additional output (or decoder) node subject to its own, individual loss function is able to well approximate the true committor function for the nucleation of methane hydrates and filter out the un-important input variables. They find that the critical variables for determination of an accurate committor function, and hence an accurate reaction coordinate, are the number of nuclear methanes, the number of cages with twelve pentagons, the number of cages with twelve pentagons and two hexagons, and the number of waters with at least three methanes within 6 \AA. Finally, the authors were able to use discovered reaction coordinate to postulate a potential nucleation mechanism.

\subsubsection{Principal Component Analysis}
At the other end of the ML complexity spectrum is the use of the aforementioned PCA technique to find the relevant order parameters for determining the crystallization phase transition in model soft matter systems. PCA can be conceptualized as a fully-connected, linear, single layer neural network with the activation function the identity function. The PCs are ordered by the percent variance of the input features they capture, and the size of the encoded or PC space is determined based on the desired mean-square reproduction of the input space on decoding. 
%\cz{since we consider PCA as ML techniques, do you think SGOOP falls into the category as well? if so, my urea from the melt paper can be cited hahaha ;)}

In a pair of papers \cite{Jadrich2018a,Jadrich2018b}, the authors demonstrate that the first PCs can be used as effective order parameters for describing phase transitions in model colloidal systems. Specifically, it is shown that the average value of the first PC can be used as a traditional order parameter for describing the generation of active particles in the RandOrg model, the isotropic-nematic and nematic-solid phase transitions in a system of hard ellipses, and percolation in the Widom-Rowlinson model. Furthermore, the variance of the first PC can be used as an effective proxy for the susceptibility at the phase transition. The use of the first PC as a proxy for the order parameter of a slow, global process in the system is reminiscent of the use of the first eigenvalue of the Markov transition matrix to describe either the slowest metastable process in a system or to approximate the committor for a two-state system. \cite{Berezhkovskii2004, Beyerle2019,Bowman2013,Buchete2008, Noe2007} Given these interesting results on model systems, it would be interesting to see this method applied to more complicated, atomistic and molecular models of soft matter systems.

Similarly, several studies have found that both PCA and VAEs are excellent methods for approximating the order parameter for phase transition in spin systems. One study has found that a VAE is able to find a one-dimensional latent space that correlates almost perfectly with the magnetization for the two-dimensional Ising model in both the ferromagnetic and antiferromagnetic cases.\cite{Wetzel2017} For the more complicated XY-model, a transformation of the two-dimensional latent space using the L$^2$ norm is required to nearly fully reproduce the canonical order parameter. It has also been shown in another study that the identification of the order parameter for the two-dimensional Ising model can be learned in an unsupervised manner by reduction to a clustering problem in the space spanned by the two highest variance PCs from PCA.\cite{Wang2016} For the more complex two-dimensional conserved order parameter Ising model, the four highest variance PCs are required due the rotational symmetry breaking in this model induced by the presence of domain wall formation in the lattice due to periodic boundaries.

\subsubsection{Spectral Gap Optimization of Order Parameters}
In a similar spirit to the variational approaches listed previously, the spectral gap optimization of order parameter (SGOOP) method is developed for a specific purpose in the identification of low dimensional OPs.\cite{Tiwary2016, Tsai2021sgoopd,Zou2023,zou2021sgoopurea} Two inputs required in this method are the estimated stationary probability distributions (typically, from long unbiased simulations or biased simulations with appropriate re-weighting scheme) and dynamical information required by the Maximum Caliber framework (from a short unbiased simulation). Its central goal is to find a large separation between visible slow modes and hidden fast processes in term of the spectral gaps of the transition probability matrix among various trial CVs. Therefore, by optimizing this objective function (i.e. the spectral gap, a similar goal to Refs. \cite{mardt2018vampnets,Scherer2019,Deuflhard2005}) with random global optimization method (e.g. basin-hopping), like all kinds of ML techniques, one is able to retrieve linear combinations of input variables at some maxima. Gas-liquid nucleation of Lennard-Jones particles,\cite{tsai2019reaction} argon, and nucleation of urea from the melt\cite{zou2021sgoopurea} are studied using the SGOOP method with typical pre-selected, physical-intuitive CVs.

\subsubsection{Other Feedforward Network Approaches}
Another successful example, which does not precisely fit into \cz{any of the three} previous categories \eb{is given in \cite{Rogal2019}, where NN-discovered RCs are coupled to an enhanced sampling technique to probe nucleation. The authors} pass symmetry functions of the atomic positions through a feedforward NN to learn a fuzzy indicator function that describes the probability of an input configuration belonging to a certain crystal phase or a disordered phase. The RC is then generated by creating a path CV in the spirit of \cite{Branduardi2007}; this path CV is biased using the driven adiabatic free energy dynamic (d-AFED) approach \cite{Abrams2008} to sample the BCC to A15 phase transition in molybdenum. Using this approach, the authors are able to describe well the layer-by-layer transition mechanism governing the nucleation of the A15 phase from the BCC phase.

\section{Machine Learning for Phase and Crystal Identification}
\label{ID}
% Start this section with a review of 'classical' methods for classifying structures and phases
% I think one weakness of the 'classical' methods is that they might struggle with 'novel'
% crystal structures (i.e. those found for specific molecules, e.g. the I form of urea), which
% is where the specificity provided by a non-linear machine learning approach can be beneficial.
\eb{In this section, we outline classic, i.e. not ML-based, methods for phase and crystal identification, with applications mostly to water and colloidal systems. Popular ML approaches for classification include graph neural networks and convolutional neural networks.}
\subsection{Classical Approaches to Phase and Crystal Identification}
Classical (non-deep) methods for classifying the physical state of a system have been in use for over thirty years, since the pioneering paper of Honeycutt and Andersen outlining the common neighbor analysis (CNA) method.\cite{Honeycutt1987} Since the publication of that paper, several other methods that do not rely on machine learning for crystal structure of systems have been developed, such as polyhedral template matching \cite{Larsen2016}, Voronoi topology analysis \cite{Lazar2015}, CHILL and CHILL+\cite{Nguyen2015} (for water only), and Ackland-Jones analysis.\cite{Ackland2006} Use of these methods has become commonplace, not in the least because they have been made available reliably and publicly through the use of freely available analysis software.

However, there are cases, in particular edge cases at the border of two phases or crystal polymorphs, where these classical methods are known to either fail completely or perform poorly at identification. In these situations, recent remedies have been found through the application of deep NNs, in particular convolutional and graph neural networks (CNNs and GNNs, respectively). NNs can be vastly more effective in correctly identifying local environments at the interfaces of crystal structures because, after training on correctly labeled samples, NNs are able to learn complex mapping functions from environment to structure label with a higher precision than the simpler classical methods. CNNs, which are widely used in classification tasks \cite{Goodfellow2016}, are particularly suited to this task of identifying crystal structures. Likewise for GNNs because the local environment of each atom or molecule in a system can be associated with a specific graph structure, leading to easy identification by GNNs. For example, the local environments for simple cubic and hexagonal close packed structures have affiliated with them do different graphs due to the differences between the units in each of those crystal structures.

\subsection{Machine-learned Approaches to Phase and Crystal Identification}

\eb{Given the increased descriptive flexibility offered by deep NNs, it should not be surprising that NN architectures of varying flavors are in continual development for the identification of both well-studied and novel phases of matter, especially in edge cases, e.g. the classification of a given species at or near a phase boundary. Two specific areas where NN classifiers have found success so far are in the classification of the phases of water-ice and the phases of several model and realistic colloidal systems.}
\subsubsection{Classifying Phases of Water}
One application of ML to ice classification. Due to its rich polymorphism and overall universal importance, the nucleation of ice crystals and general phase transitions of water are of great interest to the nucleation community. As such, several ML methods have been developed to identify the crystal structure and phase of water both in the bulk and at interfaces. 

One pioneering method utilizing deep NNs for crystal structure identification is that performed in Ref. \onlinecite{geiger2013neural}, where the authors construct Behler-Parrinello type symmetry functions for each atom in the system and use these features as input to a multilayer perceptron with a softmax output classifier. When testing this setup on Lennard-Jones systems, they find the method is more accurate than the traditional use of averaged Steinhardt bond order parameters; similar results are achieved when the machinery is used to classify ice polymorphs, especially at the ice-water interface. In both cases, the neural network identifies the correct phase with close to or greater than 90\% accuracy on the held-out test set. 
%\cz{for ice identification, should the GCICENET https://pubs.rsc.org/en/content/articlelanding/2020/cp/d0cp03456h and DeepIce https://pubs.acs.org/doi/full/10.1021/acs.jcim.9b00005 papers be mentioned? }

 A general method using only the transformed Cartesian coordinates of the waters in the simulation to perform a simple liquid-ice Ih classification is outlined in Ref. \onlinecite{fulford2019NNIce}. There, the authors perform four transformations of the Cartesian coordinates an input the concatentated output through a Geiger-Dellago neural network \cite{geiger2013neural} to obtain a two-dimensional output that classifies a given frame into liquid or ice. The authors find that the accuracy of this NN-based method is superior to the classical approaches of PTM, tetrahedral order, and Steinhardt bond order parameters. They also find that their NN is much better than using Steinhardt bond order parameters when classifying quasi-liquid water layers at the ice-water interface.

A major weakness of the ML-based classification method put forth in Ref. \onlinecite{fulford2019NNIce} is that it performs only binary classification, sorting all given inputs into either liquid or ice Ih, while it is well-known that water forms many different polymorphs, including plastic and amorphous ices.\cite{Salzmann2011}. This weakness is addressed by the graph NN approach taken in \onlinecite{Kim2020} (GCIceNet), where the developed graph NN can perform both supervised and unsupervised structure classification. The supervised approach performs the same binary classification in liquid or ice Ih as Deepice in \cite{fulford2019NNIce} with approximately the same accuracy. The unsupervised version of GCIceNet uses a graph autoencoder\cite{Kipf2016} to learn a low-dimensional latent space where the liquid water and its ice polymorphs are separable. The authors find that the unsupervised approach is vastly superior at classification than PCA or a regular autoencoder, and the most important inputs are interpreted using the relative importance technique\cite{Boattini2019}. Using just the five most important inputs still gives a model that is able to classify water's phases with greater than 95\% accuracy. Finally, the unsupervised GCIceNet latent space is able to identify quasi-liquid water at the ice-vapor interface, and interpretation of that latent space allows for the postulate that the water-vapor melting transition is heterogeneous.

Building on Refs. \onlinecite{fulford2019NNIce} and \onlinecite{Kim2020}, the graph CNN architecture developed for general phase classification in Ref.  \onlinecite{Banik2023cegann} is perhaps the most complete method developed to date for the classification of water phases, although the authors do not test its ability to separate out all the ice polymorphs from each other and liquid water. The authors of Ref.  \onlinecite{Banik2023cegann} demonstrate that their graph CNN, which they call CEGANN, is superior to SOAP\cite{De2016} and CGCNN for classifying  identifying cubic and hexagonal ice in a stacking-disordered ice lattice. They also show that CEGANN can accurately perform the binary liquid-crystal classification during the nucleation of ice I from water. However, as shown for multiple examples in \cite{Banik2023cegann}, the CEGANN architecture is transferrable to other systems, such as the measurement of zeolite nucleation, grain boundary detection, classification of space groups and carbon polymorphs, the classification of amorphous silicon, and the detection of mesophase formation in a binary mixture.

\subsubsection{Phase Classification in Colloidal and Other Soft Matter Systems}
Beyond water and ice phase behavior, using ML for phase identification has been applied to colloidal and model soft matter systems, where it is also used to study nucleation phenomenon. In \cite{Boattini2019}, the authors input averaged Steinhardt bond order parameters into an autoencoder, then perform soft clustering in the encoded space via a Gausssian mixture model (GMM) to assign frames of trajectories of hard spheres and cubes, binary mixtures, and WCA crystal growth to phases. For the hard sphere system, the performance does not appear to be significantly better than cluster in the $(\overline{q}_4,\overline{q}_6)$ space, but for all other colloidal systems, the autoencoder-GMM framework is superior to using just the averaged Steinhardt bond order parameters. However, a weakness of the study is that the results of the ML identification model are not compared to the classical methods enumerated above (PTM, CNA, VoroTop, etc.). Furthermore, the authors find that a two-dimensional encoder space is sufficient for classification of these simple soft matter systems, but the encoded dimensionality will almost certainly need to be higher if classifying more complicated (molecular) systems will a rich polymorph library.

This classification method is extended to study the fluid to AB$_{13}$ phase transition in a binary sphere system.\cite{Coli2021} In that study, the authors add hidden layers to the network presented in Ref. \onlinecite{Boattini2019} and use a more complex set of input features to the NN classifier, including all cubic and hexatic Steinhardt bond order parameters up to twelfth order as the $(\overline{q}_4,\overline{q}_6)$ is not sufficient to separate the phases for this system. Using the more complete set of input features and the more complex NN, the authors are able to classify the two types of AB$_{13}$ phases, fluid, and fcc phase with over 97\% accuracy. The classifier is then used to find the largest size of the AB$_{13}$ at all times in a seeded nucleation simulation; this information allows for a calculation of the nucleation barrier, nucleation rate, and critical supersaturation for the fluid to AB$_{13}$ transition, which allows the authors of Ref. \onlinecite{Coli2021} to postulate that the loss of mixing entropy is not the source of the high nucleation barrier for binary nucleation.

Another application to a model soft matter system comes from Ref. \onlinecite{Takahashi2021}, where the authors use MD simulations of the Gay-Berne \cite{Gay1981} and soft-core Gay-Berne \cite{Berardi2009} varieties to study the phase transition from the isotropic to nematic and smectic phases. Using their Machine Learning-aided Local Structure Analyzer (ML-LSA) \cite{Takahashi2019a} approach, the authors identified the optimal order parameter for classifying the smectic and nematic phases from a set of inputs features which are modified Steinhardt bond order parameters. \cite{Steinhardt1983} Using this machine-learned order parameter, the authors are able to postulate a three-step nucleation mechanism to transition from the isotropic to the smectic phase, with both the Gay-Berne and soft-core Gay-Berne systems converting into the nematic phase when nucleating the smectic phase from the isotropic phase.

Yet another method for classifying colloidal particles is developed in Ref. \onlinecite{OLeary2021}, where the authors develop a state classifier by determining the neighborhood graph for each particle in the colloidal system, then pass the  graph through an autoencoder, with clustering into states being performing in the autoencoder's low-dimensional latent space. The method gives the topologically correct results (correctly separating vapor and crystal phases; determining that defects occupy a large portion of crystal topology due to structural diversity), and the authors indicate their technique is more expressive and robust when classifying colloidal particles compared to the technique laid out in \cite{Boattini2019}. The authors test their classification technique on a multi-flavored colloid system and evaporation-induced colloidal self-assembly, where they deem their techniques classification results more than satisfactory.

%\cz{i still think we should have a paragraph for Mollerio's paper \cite{Banik2023cegann}, since 1) they used GAT, 2) it does a better job than classical method like CHILL+}
\section{Machine-Learned Force Fields for Nucleation}
\label{FFs}
%Here, I want to write a new subsection dealing with 1) use of classical (molecular mechanics) force fields for the simulation of nucleation processes and 2) use of DFT to get crystal structures and energies
\eb{In this section, we review how ML techniques are being used to parameterize novel force fields that bridge \emph{ab initio} accuracy with the efficiency of classical MD for studying nucleating atoms and molecules.}
\subsection{Traditional Approaches to Force Fields}
Traditionally, for the simulation of timescales beyond the nanosecond regime, electronic effects are ignored and, instead, Newtonian or Langevin dynamics are integrated at femtosecond resolution to obtain a phase-space trajectory.\cite{Cramer2013} However, for some species, such as metals, where electronic effects are crucial to accurately determining their behavior at all timescales, quantum effects must be taken into account. \eb{A significant problem} is that accurate accounting of quantum effects typically involves running computationally intensive DFT calculations at each timestep of the trajectory; this efficiency bottleneck typically limits these types of simulations to the picosecond timescale and small system sizes.\cite{Tuckerman2002}. Since nucleation is a long timescale phenomenon at the microscale and is heavily influenced by finite-size effects\cite{Honeycutt1984}, using these \emph{ab initio} MD (AIMD) techniques to simulate nucleation is not generally feasible using currently available computational resources.Thus, if one must resort to traditional \emph{ab initio} calculations, it seems the simulation of nucleation events that cannot be described accurately at the classical level is unfeasible.
\subsection{Developing Machine-learned Force Fields}
The computational strain induced by performing DFT electronic structure calculations at each step with AIMD can be eased by instead parameterizing a force field using deep NNs trained with high accuracy DFT structures. A deep NN can be trained to recognize the energy of each of these structures of a given system and, provided the training set is sufficiently diverse, this trained network can be used to accurately interpolate the energy of any other structure presented to it.\cite{Behler2015} Due to this interpolation property of the deep NN, the DFT calculations at each step can be avoided, thus making the energy calculations at each step of the trajectory significantly more efficient, allowing for the study of nucleation in systems that cannot be described accurately using classical force fields. Now we describe how different ML approaches have been applied to help model nucleation in a variety of systems.

\subsubsection{Behler-Parrinello Approaches}
Since the pioneering work of Behler and Parrinello \cite{Behler2007} to parameterize an effective force field for silicon using feedforward neural networks, many have followed suit for different atoms and molecules. Using NNs to parameterize atomic interactions is highly useful because it allows for DFT-level accuracy in the calculation of configurational energies but also allows for the determination of dynamics and thermodynamics via atomistic simulation.

At the atomic level, using NNs has been highly successful at describing the phase transitions of gallium and silicon. For gallium, the use of the Behler and Parrinello approach allows for an accurate qualitative determination of the phase diagram and physical properties (melting temperature, heat of fusion, lattice constants, and structure factor). It also gives insights into the nucleation mechanism of gallium, showing the preference for the kinetically stable $\beta$-Ga over the thermodynamically stable $\alpha$-Ga at temperatures above 174 K through the use of seeded MD simulations.

Similar techniques have been applied to develop accurate NN potentials for silicon. In Ref. \onlinecite{bonati2018solicon}, the authors train an NN potential using metadynamics-sampled configurations. They show that, coupled with an appropriate, long-ranged CV, they are able to effectively drive nucleation of silicon from the melt; however, finite-size effects are neglected. Finally, using the NN potential allows for an accurate calculation of silicon's thermodynamic (melting temperature and enthalpy and entropy of fusion) and dynamics (diffusion coefficient) at the melting point, with agreement with experiment on-par with non-NN state-of-the art methods. The same set of authors published another study a bit later \cite{Behler2008} that elaborated on the design and procedure to construct their NNs for silicon while also re-enforcing the previous study's excellent agreement with the gold standard DFT calculations, in particular the atomic energy as a function of atomic volume for multiple solid phases of silicon. 

\subsubsection{Gaussian Process Regression}
In an another approach to modelling the force field of silicon using machine learning, in Ref. \onlinecite{Bartok2018} the authors develop a ML force field using Gaussian process regression (GPR) with SOAP kernels as the basis set. \cite{Bartok2013} Using this approach, the authors are able to obtain superior agreement with the DFT values for many physical properties of silicon in both the solid and liquid forms compared to the values calculated from state-of-the-art non-ML approaches, such as DFT with tight-binding and Tersoff potentials. The authors find that their GPR technique yields better approximations of the energy per volume of various silicon polymorphs, the structure and diffusivity of liquid silicon, and several thermodynamic properties, such as the coefficient of thermal expansion and heat capacity. Finally, the GPR force field is able to model the behavior of defects quite well, in general. Weaknesses of the approach include the exclusion of at least one phase of silicon in the postulated phase diagram and the questionable transferability of the approach to elements or materials with significant long-range interactions.

Another recent example that uses the Gaussian approximation approach instead of the Behler and Parrinello approach for generating input variables to the NN parameterizing the potential is given in \cite{Kloppenburg2023} for platinum at the nanoparticle level. They find their NN-based approach is significantly better at predicting the per-atom energies compared to the embedded atom method (EAM), and they are able to calculate the elastic constants and surface energy of platinum. In addition, \cite{Kloppenburg2023} are able to perform annealing simulations of platinum nanoparticles to infer the optimal crystallization temperature for this system.

\subsubsection{Neural Network Approaches for Water Nucleation}
One rich area where the NN parameterization approach has been applied with great success has been the nucleation of ice from water. Roberto Car's group has developed a machine learned parameterization of water called the SCAN-ML potential.\cite{piaggi2022homogeneous,piaggi2021phase,zhang2021phase} The performance of the model gives nucleation rates comparable to those of experiment and is deemed to be not-worse than the atomistic TIP4P/Ice models and the coarse-grained mW model for describing the formation of ice from water. This water model has been used to refine the phase diagram of water compared to the TIP4P/Ice water model, with the experimental phase diagram used as a common reference.\cite{zhang2021phase}

In Ref. \onlinecite{Moraweitz2016}, an NN potential for water is used to discover the importance of the van der Waals interaction in determining the peculiar physical properties of water. When a correction for the van der Waals term is included in the potential, the authors are able to explain the temperature of maximum density and the negative volume of melting as well as the an improvement of the structural properties of water (i.e. radial and angular distribution functions). Finally, the authors show that the van der Waals correction is necessary to find the minimum of the total density. Thus, the message of Ref. \onlinecite{Moraweitz2016} is that the weak modulation of the van der Waals term applies important modulations to the hydrogen bond network that in turn generates some of the anomalous physical properties of water, such as the negative volume of melting.

Recently, an NN-based approach has been utilized to simulate the dissolution of NaCl crystal in water, the corollary to nucleation. \cite{ONeill2022} Interestingly, the study finds that, mechanistically, NaCl dissolution follows a very different mechanism than the nucleation of NaCl from solution, implying there is a hysteresis effect in the nucleation-dissolution process of NaCl in water. The authors find that dissolution of the NaCl crystal is rapid once a critical surface area to volume ratio is reached, and that the critical size of the NaCl cluster at which this phenomenon occurs is not equal to the critical nucleus size predicted by CNT; this qualitative mechanism is postulated to be transferrable to the dissolution of other ionic crystals in solution.

\subsubsection{Physical LassoLars}
Finally, another method that takes a tack different from the now-traditional Behler-Parrinello approach is developed in \cite{Goniakowski2022}, where the authors utilize what they call a `physical LassoLars' (PLIP) method for identifying a good potential for zinc oxide, ZnO. Like the Behler-Parrinello approach, the total potential of the system is taken to be the sum of the individual atomic energies. The force field for ZnO developed using the PLIP method  gives an accurate calculation for the phonon density of states, bulk lattice parameters, cohesive energy, radial distribution function, and surface energy for a variety of ZnO polymorphs. While the computational efficiency of the PLIP force field is inferior to classical, non-deep FFs (e.g. Tersoff or Buckingham potentials), PLIP is orders of magnitude faster than DFT with comparable accuracy, and offers a more accurate modeling of the physics of ZnO compared to the classical approaches. For example, PLIP is able to accurately predict the nucleation of the thermodynamically stable wurtzite state while use of a Buckingham potential predicts that the less stable body-centered tetragonal (BCT) crystal forms. The PLIP is also able to predict a two-step nucleation mechanism, with the wurtzite growing out of a BCT crystal, an example of the Ostwald step rule. \cite{vanSanten1984} 
% \cz{https://iopscience.iop.org/article/10.1088/1367-2630/ab4509/pdf interesting paper, using GPR to learn properties of various polymorphs}

\section{Coarse Graining through Machine Learning}
\label{CG}
\eb{Since nucleation is a rare-event problem computationally, even with accurate ML force fields for simulation, sometimes the timescales remain inaccessible. One route to accelerate sampling of nucleation events is through coarse-graining. In this section, we provide some examples of how ML is being used to effectively coarse-grain water to model its physical properties, including its ability to nucleate into a plethora of different ice polymorphs.}
\subsection{Traditional Approaches to Coarse Graining}
%mAdd some traditional CG approaches here -- Guenza, Clementi, Voth, Schweizer, etc.
Coarse-graining is one of the oldest methods for enhancing the sampling of a physical system compared to its atomistic (or complete quantum mechanical) counterpart. Traditionally, the procedure of coarse-graining involves finding and removing all degrees of freedom relaxing more quickly than the process of interest. This approach is fundamental to many physical models, such as the Born-Oppenheimer approximation in quantum mechanics,\cite{Cramer2013} the use of Mori-Zwanzig projection operators in statistical mechanics,\cite{zwanzig_book} and the many simplified (but generally successful) models for protein folding and dynamics in biophysics such as the HP and Go models.\cite{Dill1995}

Usually, finding the degrees of freedom necessary for a good coarse grain of a system involves either the development of a theory, e.g. Mori-Zwanzig projection operators\cite{zwanzig_book}, performing a Perron cluster analysis for the eigenvalues of an estimated transition operator,\cite{Noe2007,Deuflhard2005} first principles statistical mechanics theory,\cite{Schweizer2006,Guenza2018,Rudzinski2019}  knowledge gleaned from experiment, or physical and chemical intuition from experience.\cite{Amadei1993,cui2005,Zimm1956,Potoyan2013} However, recently, supervised ML has been used to find good coarse grains by penalizing discovered grains that perform poorly by failing to model the fine grained dynamics, kinetics, or thermodynamics.\cite{Husic2020,Wang2019} 

\eb{For nucleation specifically, there has been little work developing coarse graining approaches using ML techniques, with almost all applications pertaining to finding good reduced models for water. In this section, we review some methods to coarsge grain water for studying its nucleation and phase behavior.}

\subsubsection{Machine Learning Coarse-Grained Representations for Water}
One molecule for which the application of ML has had great success is water, most notably the ML-mW and ML-BOP models first put forth in \cite{Dhabal2022}. These two models are
parameterized using an approach called the hierarchical objective genetic algorithm (HOGA), which searches the parameter space of the non-bonded interactions globally using a genetic algorithm and locally using the Nelder-Mead algorithm. Taking as an interaction basis a Tersoff potential, \cite{Chan2019} find that the two models developed using this approach, the improved mW model (ML-mW) and ML-BOP, are able to better reproduce the structure and thermodynamics of water at the mesoscale level. They find this ML method is able to generate water models of increased accuracy, compared to experimental values, and are more computationally efficient with reference to traditional point-charge and polarizable water models.

Specifically, in \cite{Chan2019}, the ML-BOP model finds the correct melting temperature for water and the temperature of maximum density, although the model overestimates the diffusivity of water over a wide range of temperatures, although this shortcoming is also present for many commonly used water models such as TIP3P and mW. Furthermore, the HOGA approach is applied to refine the Stillinger-based parameters of the mW model to yield the the ML-mW, which has many improved properties compared to mW, such as improvements in the diffusivity, temperature of maximum density, and entropy and enthalpy of melting, although these improvements come by sacrificing the melting temperature, which is correctly predicted by mW but is 25 K off for ML-mW. Nucleating ice from the liquid ML-BOP, the authors are able to describe the homogeneous nucleation from supercooled ML-BOP water in detail; they find the mechanism of the nucleation follows CNT, but that the final crystal is polycrystalline, with a preference for cubic ice and an overall composition of stacking disordered ice.

The ML-BOP method is studied in more detail in \cite{Dhabal2022}, where the authors show ML-BOP's ability to reproduce the equation of state and ice-liquid thermodynamics of water. Furthermore, the authors demonstrate ML-BOP's ability to reproduce the qualitative structure of water, in particular the qualitatively correct tetrahedrality, via the radial distribution function and structure factor. Finally, it is shown that ML-BOP is superior at modelling the liquid-vapor coexistence region of water's phase diagram.

The authors show this method is transferrable to other water models by using HOGA to re-parameterize also TIP3P in \cite{Loeffler2019}; the re-parameterization yields a ML-TIP3P water model with superior predictions for the boiling point, vapor pressure, and the general behavior of water at the liquid-vapor coexistence region in general compared to TIP3P and even more highly-performing models such as TIP4P/2005. Furthermore, the nucleation rates predicted by ML-TIP3P are in far superior agreement with the measure experimental rates compared to TIP3P, TIP4P/2005, and spc/e models. Thus, in combination with the reparameterization of the mW model performed in \cite{Chan2019}, there is good evidence that the HOGA procedure is a robust method to improve water models for more accurate characterization of critical phenomena such as liquid nucleation from the vapor and nucleation of ice from supercooled water. 

The ML-BOP model has also been applied to study, in detail, the nucleation of amorphous ice structures from the liquid state. For this case, the phase transitions occur in the supercooled regime (below the glass transition temperature) and are those slow processes and hence difficult to sample using atomistic waters. In \cite{Dhabal2023}, the authors use the CG ML-BOP model to model the nucleation, including the kinetics, of the transitions of supercooled water to amorphous ice. It is shown that for low, high, and very high density amorphous ice, the ML-BOP water can effectively reproduce the structure as measured experimentally. Furthermore, through the ice nucleation simulations, the authors are able to postulate a mechanism for the phase transition of low density to high density amorphous ice via the increase of four-coordinated water molecules and multiple high-coordinated clusters, resulting in a spinodal-type phase transition. A mechanism for the formation of high-density amorphous ice from crystalline ice is also postulated, with low-density, amorphous ice playing a crucial intermediate role. 

%A successful example of ML coarse-graining applied to nucleation is demonstrated in \cite{Dhabal2022,Chan2019}, where the authors develop and apply a coarse-grained (CG) water model to the ice nucleation problem.
\section{Potential Future Research Directions for Machine Learning in Nucleation}
\label{future}
To conclude the review, here we briefly summarize what we believe to be fruitful avenues for future integration of ML techniques and nucleation studies.
\subsection{\eb{Improving the Treatment of} Finite-size Effects}
\eb{One major unsolved issue that has continuously plagued nucleation studies is the effect of finite sizes on nucleus formation.\cite{Honeycutt1984,Salvalaglio2016argon,Salvalaglio2015Faraday,Hussain2022finite-size,blow2021seven,Mahata2019finite-size} Outside of the thermodynamic limit, the simulation box size and the number of nucleating particles inside it can affect every aspect of the thermodynamics and kinetics of the nucleation process. }\cz{Moreover, solute depletion can easily affect on the formation (or growth) of critical nuclei.  A typical workaround to this challenge relies on directly increasing the size of simulation box with a trade-off in sampling efficiency. A more natural approach is the use of a constant chemical potential ensemble as developed by Karmakar \textit{et al}.\cite{karmakar2019constantmu} In addition, several approaches have been introduced to appropriately correct the thermodynamic properties \cite{salvalaglio2015molecular,Salvalaglio2016argon} and yield consistent kinetic measures \cite{Hussain2022finite-size}, where the former methods are drawn under the CNT framework and the latter requires the critical nuclei to not span over the periodic boundary.  }
%One possible way ML could help alleviate finite-size effects is through some type of reinforcement learning, where a simulation is performed in a situation close to the thermodynamics limit, where finite-size effects are negligible, and the thermodynamics and kinetic properties of that simulation are used as targets for a loss function. Then, a simulation is run at a smaller system size, and the simulation parameters are somehow updated on-the-fly until the behavior of the finite-size system behaves similarly enough to the simulation performed at the thermodynamic limit, and those simulations parameters are used to perform a production run that avoids finite-size effects. Finally, provided that the exact ML algorithm used to perform the training process is interpretable, the method for determining the factors governing the finite-size effect should be transferrable to other systems and simulation box sizes that on which the model is trained.
\subsection{Increased Use of ML for RC Discovery}
As demonstrated by the brevity of Section \ref{RCs}, there are, as of now, few published methods for automatic discovery of RCs for nucleation using ML and AI. The autoencoder framework seems nearly ideal for discovering a low dimensional RC space for nucleation, as it combines the input features in a (non)linear manner and outputs a latent space of the desired dimensionality minimizing some form of modified reconstruction error. The autoencoder loss function could be modified to drive the learned latent space to more specifically drive nucleation, e.g. a term that encourages finding RCs that maximize the average cluster size.

There are also a number of other deep learning frameworks beside the autoencoder that could be used to discover RCs. If enough information regarding the probability density along a several important OPs in the nucleated state are known, then other generative methods such as diffusing denoising probabilistic models\cite{ho2020denoising} or normalizing flows\cite{Kobyzev2020} could be used to find and sample quickly from a low dimensional latent space, accelerating the sampling of the nucleation event, even though these ML techniques do not directly introduce RCs.
\subsection{ML for Discovering RCs, Classifying Phases, and Model Development in Active Matter Systems}
%I need to survey the active matter literature more thoroughly to see what folks have actually done to
%address the issues presented for nucleation in non-equilibrium systems
All of the studies reviewed here are equilibrium in nature in that nucleation is a spontaneous process under the observed conditions, although some of these systems require enhanced sampling methods to access the post-nucleation state, whether liquid or crystal. That is, there is no external or internal energy generation in these systems. However, in the field of active matter, dynamical in addition to structural phase transitions take place. Although the order parameters and driving forces behind the phase transitions in active matter systems may differ from traditional, equilibrium nucleation events, there seems to be no reason why many of the same tools utilized above to drive and analyze nucleation simulations should not be applicable to active matter systems. 

For example, graph network approaches should still be useful for phase identification, with different input features utilized, and generative models such as VAEs should still be useful to find effective reaction coordinates and CVs for describing the \eb{equivalent of} nucleation of dynamical phases. That is, the ML techniques developed for nucleation in equilibrium systems should be completely transferrable, \emph{mutatis mutandis}, to active matter systems provided the input features to the ML models are appropriately adjusted. While ML has been use for phase identification \cite{Jeckel2019, Xue2022}, the application of ML to active matter systems appears to be focused on that and the prediction of time series and trajectories for use in reinforcement learning and navigational applications. \cite{Cichos2020} Thus, it appears to us that there is much room for ML growth in active matter, such as using ML to find interesting CG models of dynamical non-equilibrium systems and inferring the generator of the dynamics of active systems, in addition to the already mentioned use in RC and collective variable discovery.
\subsection{Increased Use of Machine-Learned Coarse-Grained Models}
So far, most applications of coarse-graining approaches coupled with ML have focused exclusively on water, at least when considering applications to the field of nucleation. As such, there is much room to enhance the marriage between coarse-graining and ML with regards to effectively modelling nucleating systems. For example, perhaps some inspiration can be taken from the success of the CG mW model that will cause researchers to seek effective CG models for other simple molecular systems using ML. For example, the Tersoff potential method is completely general and could be used to effectively coarse-grain and parameterize other simple solvent molecules similar to water, such as urea.
\subsection{Transferability of Machine-learned Potentials}
Finally, a nearly trivial problem affecting all ML methods described here is there transferability to systems beyond that on which they are parameterized. That is, more work should be done regarding how well the models built for interatomic potentials, identification of polymorphs, construction of RCs, and coarse-graining generalize beyond the training set or to similar atoms or molecules (e.g. does an RC developed for studying the nucleation of NaCl also describe well the nucleation of LiCl or KCl?) Some work on force fields developed using machine learning suggest, not surprisingly, that simpler models transfer better to unseen data \cite{Kandy2023,Benoit2021}, which is just a simple outcome of the (in)famous bias-variance tradeoff that occurs when building an approximation function for observed data.\cite{Bishop1995}
\subsection{Summary}
\eb{To conclude, here we have collected and reviewed what we hope is a representative sampling of the literature regarding the application of machine learning (ML) and artificial intelligence (AI) algorithms to the computational and theoretical study of nucleation. We have focused on four main topics (Figure \ref{fig:figure}): using ML techniques to discover good reaction coordinates for describing and driving nucleation (Sec. \ref{RCs}); the use of shallow and deep ML approaches to identify phases of matter, mostly for ice and colloidal systems (Sec. \ref{ID}) ;the use of ML techniques, mostly Behler-Parrinello type deep NNs, for the parameterization of force fields for nucleating species (Sec. \ref{FFs}); and the use of ML to help develop coarse grained water models (Sec. \ref{CG}). With the current growth rate of computing power, we anticipate that the use of ML techniques in the study of nucleation will continue to grow as well and offer insights regarding the physical mechanisms of nucleation.}
\section*{Declaration of Competing Interest}
The authors declare that they have no known competing financial interests or personal relationships that could have appeared to influence the work reported in this paper.
\section*{Data Availability}
No data were generated for the research described in this review.
\section*{Acknowledgements}
This work is entirely funded by the US Department of Energy, Ofﬁce of Science, Basic Energy Sciences, CPIMS Program, under Award DE-SC0021009.

\printcredits

%\bibliographystyle{cas-model2-names}
\bibliographystyle{unsrt}

\bibliography{refs,nucleation}

\end{document}